\documentclass[12pt]{article}

\usepackage{epsfig}
\usepackage{cite}
\usepackage{amsmath, amssymb, amsfonts}
\usepackage{color}
\usepackage{latexsym}  
\usepackage{graphicx}
\usepackage{cancel}
\usepackage[colorlinks,bookmarks]{hyperref}
\hypersetup{pdfpagemode=UseNone, pdfstartview=FitH, linkcolor=blue, citecolor=red, urlcolor=blue}

\def\be{\begin{equation}}
\def\ee{\end{equation}}
\def\ba{\begin{eqnarray}}
\def\ea{\end{eqnarray}}

\def\bdm{\begin{displaymath}}
\def\edm{\end{displaymath}}
\def\la{~\mbox{\raisebox{-.6ex}{$\stackrel{<}{\sim}$}}~}
\def\ga{~\mbox{\raisebox{-.6ex}{$\stackrel{>}{\sim}$}}~}
\def\bq{\begin{quote}}
\def\eq{\end{quote}}

 at 10truept

\newcommand{\p}{\partial}

\renewcommand{\[}{\left[}
\renewcommand{\]}{\right]}

\newcommand{\Mpl}{M_{\mathrm{Pl}}}

\newcommand{\bea}{\begin{eqnarray}}
\newcommand{\eea}{\end{eqnarray}}

\newcommand{\bi}{\begin{itemize}}
\newcommand{\ei}{\end{itemize}}

\newcommand{\beq}{\begin{equation}}
\newcommand{\eeq}{\end{equation}}
\newcommand{\beqa}{\begin{eqnarray}}
\newcommand{\eeqa}{\end{eqnarray}}
\newcommand{\mpl}{\Mpl}

\def\la{~\mbox{\raisebox{-.6ex}{$\stackrel{<}{\sim}$}}~}
\def\ga{~\mbox{\raisebox{-.6ex}{$\stackrel{>}{\sim}$}}~}

\def\12{{1 \over 2}}

\def\ltap{\ \raise.3ex\hbox{$<$\kern-.75em\lower1ex\hbox{$\sim$}}\ }
\def\gtap{\ \raise.3ex\hbox{$>$\kern-.75em\lower1ex\hbox{$\sim$}}\ }
\def\gl{\ \raise.5ex\hbox{$>$}\kern-.8em\lower.5ex\hbox{$<$}\ }
\def\roughly#1{\raise.3ex\hbox{$#1$\kern-.75em\lower1ex\hbox{$\sim$}}}

\begin{document}

\thispagestyle{empty}
\begin{flushright}
April 2025 
\end{flushright}
\vspace*{1.35cm}
\begin{center}

{\Large \bf Alternative to Axions}

\vspace*{1cm} {\large 
Nemanja Kaloper\footnote{\tt
kaloper@physics.ucdavis.edu}
}\\
\vspace{.5cm}
{\em QMAP, Department of Physics and Astronomy, University of
California}\\
\vspace{.05cm}
{\em Davis, CA 95616, USA}\\

\vspace{1.5cm} ABSTRACT
\end{center}
The $U(1)$ CP-violating phase which arises below chiral symmetry breaking 
in a non-Abelian gauge theory is ``secretly" the magnetic dual of 
the flux of a $U(1)$ $4$-form gauge field. 
Thus the discharge of this $4$-form flux by Schwinger production of charged membranes 
reduces the total CP-violating phase toward zero. This can safely 
restore CP symmetry to within the observational limits, $\theta_{\cancel{\tt CP}} \la 10^{-10}$, 
while also satisfying all limits from cosmology, 
when the charge and the tension of the gauge field sources are in the $\la keV$ range.

\vfill \setcounter{page}{0} \setcounter{footnote}{0}

\vspace{1cm}
\newpage

The observed smallness of the strong CP-violating phase $\theta_{\cancel{\tt CP}}$ is a naturalness problem 
of the Standard Model. It is a puzzle about why the phase of the quark mass matrix is so precisely aligned with
the \`a priori arbitrary vacuum angle to cancel it with the precision of $\la 10^{-10}$. The problem could be
solved dynamically, if $\theta_{\cancel{\tt CP}}$ receives a contribution from an axion, which then remains light enough
to relax total $\theta_{\cancel{\tt CP}}$ to a tiny value after QCD chiral symmetry breaking 
\cite{Peccei:1977hh,Weinberg:1977ma,Wilczek:1977pj}. Realistic models 
could be constructed, starting with \cite{Kim:1979if,Shifman:1979if,Zhitnitsky:1980tq,Dine:1981rt}.

So far, there is no direct experimental evidence for the axion, while  
limits on axion's features abound. It seems warranted to consider alternatives. 
Here we propose that $\theta_{\cancel{\tt CP}}$ is relaxed discretely, instead of smoothly,
since it includes a flux of a $4$-form (a.k.a. ``top form", since its rank equals the spacetime dimension) 
sourced by membranes, whose Schwinger discharge \cite{Schwinger:1951nm} 
rapidly relaxes $\theta_{\cancel{\tt CP}}$ right around the QCD scale.  
Our mechanism triggers a cascade of percolating bubbles 
near the QCD scale, relaxing $\theta_{\cancel{\tt CP}}$ cosmologically,
explicitly realizing some musings in \cite{Weinberg:1978uk}. This restores CP to within 
the observational limits, and  
conforms with standard cosmology since it ends well before Big Bang Nucleosynthesis (BBN). 

The presence of top forms in strongly coupled gauge theory below confinement/chiral symmetry breaking 
has been known for a long time 
\cite{Aurilia:1980xj,Aurilia:1980jz,Duff:1980qv,Luscher:1978rn,Gabadadze:1997kj,Gabadadze:2002ff,Shifman:1998if,Dvali:2003br,Dvali:2004tma,Dvali:2005an,Dvali:2005zk,Dvali:2007iv} (see 
\cite{Kaloper:2025upu} for a review). A key insight was that 
$\theta_{\cancel{\tt CP}}$ is a magnetic dual of a top form, mimicking the $U(1)$ vector gauge 
theory in $1+1$ dimension discovered in \cite{Polyakov:1975yp,DAdda:1978dle}.
In \cite{Luscher:1978rn}, the effective action of this top form
was extracted from the non-perturbative contributions to the 
correlators of the duals of gauge theory anomaly terms. Below chiral symmetry breaking, 
they have nonvanishing values in the new ground state of the theory, and hence a dynamical
top form effective action emerges. Above chiral symmetry breaking, these 
non-perturbative contributions vanish, and top form remains a total derivative, i.e. it is ``secret" \cite{Luscher:1978rn}.

The top form field equations show that it's flux is degenerate with the total CP-violating phase
$\theta_{\cancel{\tt CP}}\, $: vanishing of one implies vanishing of the other. 
In this context the axion solution of the strong CP problem is the Higgsing of the top form gauge 
theory \cite{Aurilia:1980jz,Dvali:2003br,Dvali:2004tma,Dvali:2005an,Dvali:2005zk,Dvali:2007iv}.
Even without the axion, assuming a contribution to $\theta_{\cancel{\tt CP}}$ is a magnetic dual of a 
top form, Dvali proposed a model with an enhanced density of terminal states around 
$\theta_{\cancel{\tt CP}} = 0$, making this region an attractor 
where the total $\theta_{\cancel{\tt CP}}$ can relax 
by a discharge of membranes charged under this $4$-form \cite{Dvali:2005zk}. 
This can sidestep the concerns
that nucleation rates of membranes ``native" to QCD are too slow \cite{Shifman:1998if,Forbes:2000et}
(counterexamples to these ideas are offered in \cite{Dvali:1998ms,Dubovsky:2011tu}). 

Another way to sidestep the question of whether nucleation rates of QCD membranes are fast enough
to discharge  $\theta_{\cancel{\tt CP}}$, which we pursue here,  
is to invoke a new sector whose membranes have \underbar{smaller}  
charge and tension, which are weakly coupled to QCD. To this end, 
we use a replica of L\"uscher's non-perturbative top form in QCD, but add membranes 
with tensions and charges in the $keV$ range. The new $4$-form is exactly degenerate 
away from the membranes with L\"uscher's top form, and so its discharge by membrane 
nucleation also reduces the strong
CP-violating phase $\theta_{\cancel{\tt CP}}$. This occurs near the QCD scale, when 
the background universe is radiation dominated, and expanding slowly. Hence 
the bubbles of true vacuum bounded by membranes will expand at the speed of light, 
colliding with each other and easily percolating \cite{Guth:1982pn,Turner:1992tz,Freese:2004vs}. 
The colliding membranes decay into the strongly-coupled 
QCD excitations. The vacuum with a tiny $\theta_{\cancel{\tt CP}}$ 
is achieved quickly over large regions of the universe. 
Working in the limit $\mpl \rightarrow \infty$, we ignore any interplay of the strong 
CP problem with the cosmological constant, fine tuning the latter away. We also set aside 
questions of UV completion of the mechanism at this time.

Conceptually, our mechanism is actually very simple. As is well known, the strong CP violation is encoded by 
an arbitrary $U(1)$ phase of the theory, which takes any value on 
$S^1$. The theory has an intrinsic higher rank $U(1)$ gauge field in the strong coupling regime, revealed
by computing non-perturbative corrections below chiral symmetry breaking, by e.g. evaluating
the partition function and the correlators of the anomaly operator 
$q(x) = \frac{g^2}{64\pi^2} \epsilon^{\mu\nu\lambda\sigma} F_{\mu\nu\lambda\sigma}$ \cite{Luscher:1978rn}. 
Above chiral symmetry breaking 
these correlators are zero; below, they do not vanish as shown by L\"uscher. The operator
$q$ is the dual of the emergent $4$-form field strength $F_{\mu\nu\lambda\sigma}$. 
Its flux is dual to the total CP-violating phase \cite{Luscher:1978rn,Shifman:1998if,Dvali:2005zk}. 

When we kinetically mix this $4$-form to an  additional external $4$-form gauge field with 
a sufficiently light charge, the flux will discharge by quantum nucleations of charges, as shown by
Schwinger \cite{Schwinger:1951nm}. This can be fast enough to restore CP before BBN, and so all 
of tested cosmology can remain unaffected; the mechanism can pass both the particle physics and
cosmological tests at the same time. The charge carriers are not point particles
but membranes; this is merely a technical nuisance, which is understood well enough by now.
In fact, on spherically symmetric backgrounds, when we integrate out the angular coordinates, the resulting
picture is identical to the discharge of a uniform electric field inside a parallel plate capacitor!
The top form sector which relaxes CP is a generalization to three spatial dimensions
of a Maxwell $U(1)$ gauge theory in one dimension of space. One can see this by noting 
that for well separated spherical membranes one can integrate angular coordinates and reduce
the theory to $1+1$ dimensions. The membranes, being codimension-1 objects, 
correspond to points on a line. The discharges are but particle production in a background electric
field, akin to how a parallel plate capacitor discharges by pair production. 

Let us first formulate the strong CP problem using top forms. The perturbative definition 
of the gauge sector of QCD must be extended by an extra topological term, which follows 
because of the nontrivial topology of SU(3): 
\be
{\cal L}_{\tt QCD} \ni \frac{g^2}{64\pi^2} \theta \epsilon^{\mu\nu\lambda\sigma}  
\sum_a G^a_{\mu\nu} G^a_{\lambda\sigma} \, , 
\label{cpterm}
\ee
with a canonically normalized gauge field strength 
$G^a \in SU(3)$. We sum over colors, with $\theta$ a ``vacuum angle", $\theta \in [0,2\pi]$. Aside from the 
$1/g^2$ factor in the gauge kinetic term, we follow the normalizations in \cite{Luscher:1978rn}.
This term ensures invariance of the theory under large gauge transformations; $\theta$ 
is \`a priori arbitrary; loosely, a nonzero $\theta$ implies CP violation. 

However, with unbroken chiral symmetry, all values of $\theta$ are degenerate. 
To see it, we can rewrite (\ref{cpterm}) using $\frac{g^2}{32} \epsilon^{\mu\nu\lambda\sigma} 
\sum_a G^a_{\mu\nu} G^a_{\lambda\sigma} = \partial_\mu K^\mu$, 
where $K_\mu$ is the Chern-Simons current. Hodge-dualizing, 
$K_\mu = \epsilon_{\mu\nu\lambda\sigma} A^{\nu\lambda\sigma}/6 $, brings about the
first apparition of QCD's ``secret" top form  \cite{Luscher:1978rn}: 
$\partial_\mu K^\mu = \frac16 \epsilon^{\mu\nu\lambda\sigma} \partial_\mu A_{\nu\lambda\sigma} = 
\frac{1}{24} 
\epsilon^{\mu\nu\lambda\sigma} F_{\mu\nu\lambda\sigma}$ with 
$F_{\mu\nu\lambda\sigma} = 4  \partial_{[\mu} A_{\nu\lambda\sigma]}$. 
This yields $F_{\mu\nu\lambda\sigma} = \frac34 g^2 \sum_a G^a_{[\mu\nu} G^a_{\lambda\sigma]}$, 
where $[\ldots]$ denotes antisymmetrization. So the $\theta$ term in the perturbative ${\cal L}_{\tt QCD}$ is 
$\frac{1}{48 \pi^2} \theta \epsilon_{\mu\nu\lambda\sigma} F^{\mu\nu\lambda\sigma}$, which is a total derivative. 

With massive quarks, whose mass matrix is ${\cal M}$, an additional contribution comes from  
${\tt Arg} \det {\cal M}$. In the fermion basis where $\det {\cal M}$ is real, the total CP-violating phase is 
$\theta \rightarrow \theta_{\cancel{\tt CP}} = \hat \theta = \theta + {\tt Arg} \det {\cal M}$. 
Even so, different $\theta_{\cancel{\tt CP}}$'s
are still degenerate above the QCD scale, before the instanton effects kick in. 
Further, these superselection sectors disappear if even one of the quarks were massless, since then the 
determinant of ${\cal M}$ would have been zero, and so ${\tt Arg} \det {\cal M}$ would
have been completely arbitrary. If so, we could cancel any value of $\hat \theta$ 
by picking at will an appropriate value of ${\tt Arg} \det {\cal M}$.

Now we turn to the top form $F^{\mu\nu\lambda\sigma}$ below chiral symmetry breaking. 
Following \cite{Luscher:1978rn} the operator expectation values are 
\be
\langle {\cal O} \rangle_{\hat \theta} = \frac{1}{{\cal Z}[{\hat \theta}]} 
\int \[{\cal D}{\cal B}\] {\cal O} \, e^{i S + i \hat \theta \int q(x)} \, , ~~~~~~ 
{\cal Z}[{\hat \theta}] =  \int \[{\cal D}{\cal B}\] \, e^{i S + i \hat \theta \int q(x)} \, ,
\label{partition}
\ee
where $q(x) = \frac{g^2}{64\pi^2} \epsilon^{\mu\nu\lambda\sigma} 
\sum_a G^a_{\mu\nu}(x) G^a_{\lambda\sigma}(x) 
= \frac{1}{48 \pi^2} \epsilon_{\mu\nu\lambda\sigma} F^{\mu\nu\lambda\sigma}$, 
as explained above. 
In (\ref{partition}) we decoupled the fermions for simplicity, by assuming they are massive, which 
absorbs the mass matrix $U(1)$ phase into $\hat \theta$, and leaves us 
with a path integral over the gauge fields $B^a_\mu = \{{\cal B}\}$. Although the quark mass 
scale is comparable to the chiral symmetry breaking scale in real QCD, 
this simplification captures the key features of the dynamics well. The Euclidean 
partition function ${\cal Z}[{\hat \theta}]$ 
has minima at $\hat \theta =0 + 2n\pi$ \cite{Vafa:1984xg}, where CP is restored, and so 
it is an even function of $\hat \theta$ \cite{Vafa:1984xg}. 

As \cite{Luscher:1978rn}, we are interested in the correlators of $q(x)$. \underbar{Below chiral symmetry breaking},
they can be computed using instanton dilute gas approximation. First off, $\langle q \rangle_{\hat \theta}$ is not
vanishing unless $\hat \theta = 0$:
\be
\langle q \rangle_{\hat \theta} = \frac{1}{{\cal Z}[{\hat \theta}]} 
\int \[{\cal D}{\cal B}\] q(x) \, e^{i S + i \hat \theta \int q(x)} = i {\tt H}(\hat \theta) \, , \label{q}
\ee
where ${\tt H}(- \hat \theta) = - {\tt H}(\hat \theta)$. For $SU(2)$ in a particular $\hat \theta$ state,  
L\"uscher found this function to be
${\tt H}(\hat \theta) = 0.078 (\mu_0 \frac{8\pi^2}{g^2})^4 (\mu_0 \rho_c)^{10/3} e^{-8\pi^2/g^2(\mu_0)} \sin(\hat \theta)$,
where $\mu_0$ is a dimensional normalization parameter, $g(\mu_0)$ the running coupling and
$\rho_c$ the IR cutoff for the instanton size. A qualitatively similar result also holds for $CP^{n-1}$ 
nonlinear $\sigma$-models in $1+1$ dimensions. Similar features are also found in more realistic setups in 
\cite{Gabadadze:1997kj,Shifman:1998if,Gabadadze:2002ff}. Note that because 
the vacuum angle $\hat \theta$ is  
removable by chiral transformations \underbar{above complete breaking of chiral symmetry}, where at least one fermion remains massless, this expectation value must be zero at scales above $\Lambda_{\tt QCD}$. 
Thus, since as noted above $q(x) = \frac{1}{48 \pi^2} \epsilon_{\mu\nu\lambda\sigma} F^{\mu\nu\lambda\sigma}$, 
below chiral symmetry breaking
the ``secret" top form field strength is not zero when $\hat \theta$ is not zero since $q \ne 0$: 
inverting the Hodge dual, and transforming to Lorentzian 
signature variables, 
\be
F_{\mu\nu\lambda\sigma} = - 2\pi^2 i \epsilon_{\mu\nu\lambda\sigma} q 
= 2\pi^2 \epsilon_{\mu\nu\lambda\sigma} {\tt H}(\hat \theta) \, . \label{Ftheta}
\ee
Hence $F_{\mu\nu\lambda\sigma} \ne 0$ is a faithful diagnostic of CP violation parameterized by
$\hat \theta$. Further, restoring CP is equivalent to selecting a state in which the top form
field strength is zero. 

There are also additional  $F^{\mu\nu\lambda\sigma}$-dependent corrections \cite{Luscher:1978rn}. 
They are duals to the higher-order correlators $\langle q(x_1) q(x_2) \ldots q(x_n) \rangle_{\hat \theta}$,
which are nonzero below chiral symmetry breaking/confinement scale, 
while vanishing when chiral symmetry is restored. 
For our purposes it suffices to keep only quadratic terms, and ignore higher powers. To extract these terms
from the partition function in the IR, following 
\cite{Luscher:1978rn,Gabadadze:1997kj,Shifman:1998if,Gabadadze:2002ff}, we can define a gauge field
${\cal Y}_\mu = \epsilon_{\mu\nu\lambda\sigma} A^{\nu\lambda\sigma}/6$ such that 
$q = \frac{1}{2\pi^2} \partial_\mu {\cal Y}^\mu$, compute the $1$-PI correlator 
$\langle q(x) q(y) \rangle_{\hat \theta}$ and, using Lorentz symmetry, extract from it the correlator for
$\langle {\cal Y}_\mu {\cal Y}_\nu \rangle_{\hat \theta}$. After some straightforward albeit tedious 
calculations, using the defining relation of ${\cal Y}$ in terms of $A$ above, we finally find 
that in Lorenz gauge $\partial_\mu A^{\mu\nu\lambda} = 0$ the two-point function for
the top form field potential is 
\be
\frac{1}{(3!)^2} \epsilon_{\mu\alpha\beta\gamma} \epsilon_{\nu\sigma\rho\delta}  
\langle A^{\alpha\beta\gamma}(x) A^{\sigma\rho\delta}(y)\rangle_{\hat \theta} 
= 4\pi^4 \int \frac{d^4p}{(2\pi)^4} e^{ip(x-y)} \frac{p_\mu p_\nu}{p^4}  {\cal X} \, , 
\label{greens}
\ee
where ${\cal X} =  - i \int d^4x \, \langle T\bigl(q(x)q(0)\bigr) \rangle_{\hat \theta} = \frac{d}{d\hat \theta} {\tt H}(\hat \theta) \simeq (\Lambda_{\tt QCD})^4$ is the topological susceptibility 
of the theory. The relationship 
${\cal X} =  \frac{d}{d\hat \theta} {\tt H}(\hat \theta)$ for momentum space correlator in Eq. (\ref{greens}) 
follows directly from 
$\int d^4y \langle q(x) q(y) \rangle_{\hat \theta} = \frac{1}{i} \frac{d}{d\hat \theta} \langle q \rangle_{\hat \theta}$
and Eq. (\ref{q}), after Fourier transforming to momentum space and using 
Lorentz invariance of the ground state: it shows that
the nonperturbatively induced powers of $F_{\mu\nu\lambda\sigma}$ have correlated coefficients. Hence well below 
$\Lambda_{\tt QCD}$, the effective theory can be thought of as an operator 
series including terms  $\propto \bigl(F_{\mu\nu\lambda\sigma}\bigr)^n$.

From Eq. (\ref{greens}), we confirm that 1) above the complete chiral symmetry symmetry breaking scale, when at
least one quark is massless, ${\tt H} = {\cal X} = 0$, the QCD top form is ``secret": the term linear 
in flux is physically 
irrelevant since we can pick $\hat \theta$ to be zero by a chiral gauge 
transformation, and the higher power correlators, 
starting with the two-point function $\langle A A \rangle_{\hat \theta}$, do not appear in the effective
action, and 2) below chiral symmetry breaking, when ${\cal X} \ne 0$, and so ${\tt H}$ is a non-trivial function of
$\hat \theta$, 
the effective theory must include terms $\propto \bigl( F_{\mu\nu\lambda\sigma}^2 \bigr)^{n}$, starting with 
\be
{\cal L}_{\tt QCD} \ni \frac{1}{48 \pi^2} 
\hat \theta \epsilon_{\mu\nu\lambda\sigma} {F}^{\mu\nu\lambda\sigma}  
-  \frac{1}{4 \pi^2 \cdot 4! \pi^2 {\cal X}} {F}_{\mu\nu\lambda\sigma}^2  \, ,
\label{efflag} 
\ee
up to gauge-fixing terms, that reproduces the correlator in Eq. (\ref{greens}). Recall, that classically the $4$-form field strength does not include 
new local degrees of freedom. For a purely quadratic 
theory $\epsilon^{\mu\nu\lambda\sigma} {F}_{\mu\nu\lambda\sigma} = {\rm const}$ by equations
of motion, which is the hidden constant elucidated in \cite{Aurilia:1980xj}. 

It is convenient to replace the field strength ${F}_{\mu\nu\lambda\sigma}$ with its magnetic dual
$\propto {F}_{\mu\nu\lambda\sigma} \epsilon^{\mu\nu\lambda\sigma}/4!$. This is the field
theoretic realization of the canonical transformation exchanging momenta and coordinates. 
The procedure is discussed at length in e.g. 
\cite{Aurilia:1978dw,Aurilia:1980xj, Dvali:2005an,Kaloper:2008qs,Kaloper:2008fb,Kaloper:2011jz,Kaloper:2016fbr,Kaloper:2022oqv,Kaloper:2022utc}. 
Here we will merely summarize the main steps. We start with the action based on the Lagrangian (\ref{efflag}). 
Recalling that 
$F_{\mu\nu\lambda\sigma} = 4 \partial_{[\mu} A_{\nu\lambda\sigma]}$, we rewrite this action in the
first order formalism, employing a Lagrange multiplier $F$: 
\be
S = \int d^4 x \Bigl(- \frac{1}{4\pi^2 \cdot 4! \pi^2 {\cal X}} { F}_{\mu\nu\lambda\sigma}^2 
+  \frac{\hat \theta}{2 \cdot 4! \pi^2} 
\epsilon_{\mu\nu\lambda\sigma} { F}^{\mu\nu\lambda\sigma} + 
\frac{F}{4 \pi^2 \cdot 4! \pi^2 {\cal X}} 
\epsilon_{\mu\nu\lambda\sigma} \bigl( { F}^{\mu\nu\lambda\sigma} -  4 \partial_{\mu} A_{\nu\lambda\sigma} \bigr) \Bigr)  \, .
\label{cantraax1}
\ee
Next we complete the square for ${ F}_{\mu\nu\lambda\sigma}$, by defining
$\tilde { F}_{\mu\nu\lambda\sigma} = { F}_{\mu\nu\lambda\sigma} - (F+2\pi^2 {\cal X} \hat \theta) \epsilon_{\mu\nu\lambda\sigma}$, and integrate $\tilde { F}_{\mu\nu\lambda\sigma}$ out. This only yields a
Gaussian factor in the partition function since the Jacobian is unity. 
The precise details of this calculation are given in \cite{Kaloper:2008qs,Kaloper:2008fb,Kaloper:2011jz,Kaloper:2016fbr,Kaloper:2022oqv,Kaloper:2022utc} and we
do not repeat them here. After integrating out $\tilde { F}_{\mu\nu\lambda\sigma}$, the remaining terms are 
\be
S = \int d^4 x \Bigl(- \frac{{\cal X}}{2}  \bigl(\hat \theta 
+ \frac{F}{2\pi^2 {\cal X}} \bigr)^2 - \frac{F}{24 \pi^4 {\cal X}}  
\epsilon^{\mu\nu\lambda\sigma}  \partial_\mu {A}_{\nu\lambda\sigma} \Bigr) \, .
\label{cantrad}
\ee
The variation with respect to ${A}_{\nu\lambda\sigma}$ yields field equation for $F$: $\partial_\mu F = 0$. 
Thus the local solutions for the field $F$ are just constants. The quadratic term 
\be
V = \frac{{\cal X}}{2}  \bigl(\hat \theta + \frac{F}{2\pi^2 {\cal X}} \bigr)^2 \, , 
\label{potential}
\ee
is the contribution to the vacuum energy when CP is violated, calculated in
e.g. \cite{Gabadadze:1997kj,Gabadadze:2002ff}. This shows that the total CP-violating phase
actually is 
$\theta_{\cancel{\tt CP}} = \hat \theta + \frac{F}{2\pi^2 {\cal X}}$. It sources the field 
${F}_{\mu\nu\lambda\sigma} = 4 \partial_{[\mu} A_{\nu\lambda\sigma]}$, 
as can be seen varying (\ref{cantrad}) with respect to $F$; 
after some manipulation \cite{Kaloper:2025upu}, ${F}_{\mu\nu\lambda\sigma} 
= 2\pi^2 {\cal X}\theta_{\cancel{\tt CP}} \, \epsilon_{\mu\nu\lambda\sigma}$. 
So $F_{\mu\nu\lambda\sigma}$ probes the presence of total CP violation: 
if $\theta_{\cancel{\tt CP}}$ vanishes, so does $F_{\mu\nu\lambda\sigma}$ (but not the dual $F$; this 
feature is a well known property of Legendre transforms). 
Next, although $F$ is locally constant, because it is a magnetic 
dual of $F_{\mu\nu\lambda\sigma}$, it can vary from region to region when there are charges 
sourcing it \cite{Gnadig:1976pn,Luscher:1978rn,Aurilia:1978dw,Gabadadze:1997kj,Gabadadze:2002ff}. 
The mechanics of quantum discharges is discussed at length in, e.g. \cite{Brown:1987dd,Brown:1988kg}. 

The discharges might be very 
slow \cite{Shifman:1998if,Forbes:2000et}, but there are examples which suggest otherwise 
\cite{Dvali:1998ms,Dubovsky:2011tu}. We sidestep this, and 
add one more copy of the QCD top form sector. We also add membranes charged under it, 
with tension ${\cal T}$ and charge ${\cal Q}$, treating them 
as the input parameters of the theory. As we are interested in the proof-of-principle, we pick 
the scales ${\cal T}, {\cal Q}$ such that the relaxation processes are able to resolve the strong CP problem.
It turns out that a tension and charge $\sim keV$ do the job. 

To include our new top form, we imitate the axion 
couplings \cite{Weinberg:1977ma,Wilczek:1977pj} (see \cite{Kaloper:2025upu} for details): 
\ba
S_{{F} + {\cal H}} &=& \int d^4x \Bigl(- \frac{1}{2}  \bigl( {\cal H} 
+\sqrt{\cal X} \hat \theta+  \frac{F}{2\pi^2 \sqrt{\cal X}} \bigr)^2+\frac{\epsilon^{\mu\nu\lambda\sigma}}{24\pi^4 {\cal X}} 
  \partial_\mu \bigl({ F }\bigr) {A}_{\nu\lambda\sigma} 
 + \frac{\epsilon^{\mu\nu\lambda\sigma}}{6}  
 \partial_\mu \bigl( {\cal H} \bigr) {\cal C}_{\nu\lambda\sigma} 
\Bigr) \nonumber \\
&-& {\cal T} \int d^3 \, \xi \sqrt{|\det(\eta_{\mu\nu} \frac{\p x^\mu}{\p \xi^a} \frac{\p x^\nu}{\p \xi^b} )|} 
- \frac{\cal Q}{6} \int d^3 \xi \, {\cal C}_{\mu\nu\lambda} \frac{\p x^\mu}{\p \xi^a} \frac{\p x^\nu}{\p \xi^b} 
\frac{\p x^\lambda}{\p \xi^c} \epsilon^{abc}  \, , ~~~~
\label{cantradcharged}
\ea
where ${\cal H}$ is the dual of the new top form\footnote{The bilinears $\propto \epsilon^{\mu\nu\lambda\sigma}$ 
were integrated by parts to covariantize the top forms on the membranes \cite{Kaloper:2022oqv,Kaloper:2022utc}.}.  
Varying (\ref{cantradcharged}) yields $\frac{1}{2\pi^2 \sqrt{\cal X}} {F}_{\mu\nu\lambda\sigma} 
= {\cal H}_{\mu\nu\lambda\sigma} 
= \Bigl( {\cal H} + \sqrt{\cal X}  \hat \theta +  \frac{F}{2\pi^2 \sqrt{\cal X}} \Bigr) \, \epsilon_{\mu\nu\lambda\sigma}$. 
As before, away from the charges both fluxes $F$ and ${\cal H}$ are constant. Once
charges are included, the flux ${\cal H}$ can change discretely.
Note that the CP-violating phase now is 
$\theta_{\cancel{\tt CP}} =  \frac{\cal H}{\sqrt{\cal X}} + \hat \theta + \frac{F}{{2\pi^2 {\cal X}} }$. 
The $3$-form ${\cal C}$ comes in as a Lagrange multiplier enforcing $\partial_\mu {\cal H}=0$. It yields   
${\cal H}_{\mu\nu\lambda\sigma} = 4 \partial_{[\mu} {\cal C}_{\nu\lambda\sigma]}$ on shell. Note that if we integrate out 
the angular variables on $S^2$ and fix $F$, Eq. (\ref{cantradcharged}) is just the standard Maxwell theory in 
one time and one spatial dimension, written in terms of the dual magnetic field. 

To determine the quantum nucleation rates, we euclideanize this action, working in the decoupling limit of gravity 
$\mpl \rightarrow \infty$ \cite{Coleman:1977py}. 
Following \cite{Coleman:1977py,Callan:1977pt,Garriga:1993fh} and \cite{Kaloper:2022oqv,Kaloper:2022utc}, 
the Wick-rotated Euclidean action defined by $i S \rightarrow - S_E$ is 
\ba
S_E &=&\int d^4x \Bigl( \frac{1}{2}  \bigl( {\cal H} +\sqrt{\cal X} \hat \theta+ \frac{F}{2\pi^2 \sqrt{\cal X}}   \bigr)^2 
+  \frac{{\epsilon^{\mu\nu\lambda\sigma}} }{24\pi^4 {\cal X}} 
\partial_\mu \bigl( {F} \bigr) { A}_{\nu\lambda\sigma} 
+ \frac{{\epsilon^{\mu\nu\lambda\sigma}}}{6}  \partial_\mu \bigl( {\cal H} \bigr) {\cal C}_{\nu\lambda\sigma} \Bigr)  
\nonumber \\
\label{actionnewmemeu}
&+& {\cal T} \int d^3 \xi \sqrt{\gamma}_{\cal C} - \frac{{\cal Q}}{6} \int d^3 \xi \, {\cal C}_{\mu\nu\lambda} \, 
\frac{\p x^\mu}{\p \xi^\alpha} \frac{\p x^\nu}{\p \xi^\beta} 
\frac{\p x^\lambda}{\p \xi^\gamma} \epsilon^{\alpha\beta\gamma} \, .
\ea
When $\mpl \rightarrow \infty$ the only available nucleation channel is the simplest bounce solution 
relating the backgrounds with nonnegative vacuum energy 
$V = \frac{1}{2}  \bigl( {\cal H} +\sqrt{\cal X} \hat \theta+  \frac{F}{2\pi^2 \sqrt{\cal X}} \bigr)^2 $ 
where  ${\cal H}_{in,out}$ differ by a unit of charge, ${\cal H}_{out} - {\cal H}_{in} = {\cal Q}$. 
Although ${\cal H}$ is quantized, ${\cal H} = {\cal N} {\cal Q}$, it is 
degenerate with $\sqrt{\cal X} \hat \theta$ and  $\frac{F}{2\pi^2 \sqrt{\cal X}}$, which  
can take arbitrary values when there are no restrictions on the overall phase of the 
quark mass matrix, and so we ignore the ${\cal H}$ quantization. 

A membrane which nucleates must also satisfy energy conservation: 
the energy difference in the interior must equal the 
rest energy set by the tension \cite{Coleman:1977py}. 
Subtracting the backgrounds with and without a membrane yields 
the membrane action $S_{membrane} = 2\pi^2 r_0^3 {\cal T} - \frac12 \pi^2 r_0^4 \Delta V$, 
that follows since the volume 
integrals are $V_{S^4} = \pi^2 r_0^4/2$ and $V_{S^3} = 2\pi^2 r_0^3$. 
The bounce is the minimum with respect to $r_0$ 
which yields $r_0 = {3{\cal T}}/{\Delta V }$. The actual bounce action is \cite{Coleman:1977py}
\be
B = \frac{27\pi^2}{2} \frac{{\cal T}^4}{\bigl(\Delta V\bigr)^3} \, .
\label{bounce}
\ee

The bubble nucleation rate per unit time per unit volume is 
$\Gamma = A e^{-B}$ \cite{Coleman:1977py,Callan:1977pt}. The prefactor $A$ for membrane production 
in flat space was found by Garriga in 
\cite{Garriga:1993fh}, and so  
\be
\Gamma \simeq 9 \frac{{\cal T}^4}{\bigl(\Delta V\bigr)^2} 
\exp\Bigl({-\frac{27\pi^2}{2} \frac{{\cal T}^4}{\bigl(\Delta V\bigr)^3}}\Bigr) \, .
\label{nucrate}
\ee
Inside a membrane, the potential difference is 
$\Delta V \simeq  {\cal Q}  \bigl( {\cal H} +\sqrt{\cal X} \hat \theta+  \frac{F}{2\pi^2 \sqrt{\cal X}} \bigr)$, 
which receives ${\cal O}(1)$ corrections when the QCD top form is near zero. 
The QCD top form also decreases by 
$\Delta { F}_{\mu\nu\lambda\sigma} = {2\pi^2 \sqrt{\cal X}}  {\cal Q} \, \epsilon_{\mu\nu\lambda\sigma}$, since
both it and the potential are set by $\theta_{\cancel{\tt CP}}$. Since the membrane tension is strictly positive,
discharges decrease the potential and 
$\theta_{\cancel{\tt CP}}$ until $| F_{\mu\nu\lambda\sigma} | <{2\pi^2 \sqrt{\cal X}}  {\cal Q}$: 
since ${\cal T} > 0$, and
$r_0 = 3 {\cal T}/\Delta V$, when $\Delta V <0$ we'd need $r_0<0$, which is clearly unphysical. 

With a single membrane, the requirement $\theta_{\cancel{\tt CP}} \la 10^{-10}$ and the bound 
$| F_{\mu\nu\lambda\sigma} | < {2\pi^2 \sqrt{\cal X}}  {\cal Q}$ impose
$ {{\cal Q}}/{\sqrt{\cal X}} \la 10^{-10}$. 
Using ${\cal X}^{1/4} \simeq {\rm few} \times 100 MeV$, we find 
$ {\cal Q} \la 10^{-11} GeV^2 \simeq (3 keV)^2$. Further, the 
discharge mechanism should be faster than the cosmic dilution close to the QCD scale, 
$\Gamma \ga H_{\tt QCD}^4$ \cite{Guth:1982pn,Turner:1992tz,Freese:2004vs}. 
Since $H_{\tt QCD} \simeq (100 MeV)^2/\mpl \simeq 10^{-20} GeV$, we require 
\be
10^{81} \frac{{\cal T}^4}{\bigl(\Delta V\bigr)^2(GeV)^4} \ga 
\exp\Bigl({\frac{27\pi^2}{2} \frac{{\cal T}^4}{\bigl(\Delta V\bigr)^3}}\Bigr) \, .
\label{boundrate}
\ee
The nucleations are the slowest close to the terminal 
state \cite{Kaloper:2025upu}, where $\Delta V$ is the smallest; there 
$\Delta V = \bigl( {\cal Q} \bigr)^2 \la 10^{-22} GeV^4$. The decay rate will be fast enough, $\Gamma > H_{\tt QCD}^4$, 
if ${\cal T}^{1/3} \la 3 keV$. Note that since the decay rate depends on the tension and charge exponentially, even if 
the actual value of the strong CP phase $\theta_{\cancel{\tt CP}}$ is a few orders of magnitude smaller than $10^{-10}$,
which could be tested by future experiments \cite{Anastassopoulos:2015ura,Zhevlakov:2020bvr}, 
our mechanism would remain
operational. For example, if $\theta_{\cancel{\tt CP}}$ were as small as $10^{-13}$, the relaxation of 
$\theta_{\cancel{\tt CP}} \sim {\cal O}(1)$ to $10^{-13}$ would still occur if the charge and tension were
$\sim {\cal O}(100) \, eV$ (see \cite{Kaloper:2025upu} for details). 

We can now confirm that neglecting gravity is justified. Since $H_{\tt QCD} \simeq \sqrt{\cal X}/\mpl$, 
the radius of a membrane at nucleation is 
$r_0 H_{\tt QCD} \simeq \frac{\cal T}{\Delta V} \frac{\sqrt{\cal X}}{\mpl} 
\simeq \frac{\cal T}{\sqrt{\cal X}\mpl} \frac{1}{\theta_{\cancel{\tt CP}} \Delta \theta_{\cancel{\tt CP}}}$. 
For transitions close to the final state, with $\theta_{\cancel{\tt CP}} \la 10^{-10}$, 
this means $r_0 H_{\tt QCD} \la 10^{-12}$. 
Initially, for $\theta_{\cancel{\tt CP}} \sim 1$, the bubbles are even much smaller, with $r_0 H_{\tt QCD} \la 10^{-22}$. 
Hence our bubbles are very small compared to the background 
curvature scale $1/H_{\tt QCD} \simeq \mpl/\sqrt{\cal X} \simeq 10^{11} (eV)^{-1} \simeq 100 \, km$, 
and we can always pick a large freely falling local frame, smaller than $1/H_{\tt QCD}$ and much 
larger than the size of a nucleating bubble. 

Now we turn to the cosmology of CP restoration. We start with 
the hot universe after inflation, evolving as a radiation-dominated, spatially flat, 
isotropic and homogeneous FRW cosmology, with temperature $T$ 
well above $\Lambda_{\tt QCD} \sim GeV$. As long as 
$T \gg \Lambda_{\tt QCD}$, the QCD chiral symmetry will not be broken, and so, initially, all the 
$\theta_{\cancel{\tt CP}}$ states will be degenerate, due to massless quarks. There will be no
QCD-related CP violation; but as the universe cools, electroweak symmetry breaking at $T \sim TeV$ will fix 
the quark mass matrix phase contribution to $\theta_{\cancel{\tt CP}}$, initiating chiral symmetry breaking, 
and subsequent contributions to chiral symmetry breaking will come 
when the instanton effects arise, which will lift the degeneracy of  $\theta_{\cancel{\tt CP}}$. 
Concurrently, the discharges of $\theta_{\cancel{\tt CP}}$ induced 
by discharges of ${\cal H}$ will remain dormant until $H \sim H_{\tt QCD}$ 
because the vacua with different $\theta_{\cancel{\tt CP}}$ were initially degenerate, and the
QCD top form and ${\cal H}$ are decoupled before the QCD scale. 
The instanton effects kick in around the QCD scale, explicitly lifting the degeneracy of different 
$\theta_{\cancel{\tt CP}}$ states. The domains with a value 
of $\theta_{\cancel{\tt CP}}$ develop vacuum energy
$V = \frac12 {\cal X} \theta_{\cancel{\tt CP}}^2 = \frac{1}{2}  
\bigl( {\cal H} +\sqrt{\cal X} \hat \theta+  \frac{F}{2\pi^2 \sqrt{\cal X}} \bigr)^2$. 
Those regions are unstable to rapid nucleation of small bubbles surrounded by 
membranes with tension ${\cal T}^{1/3} \sim 3 keV$ and charge $\sqrt{{\cal Q}} \sim 3 keV$. 
They will decay toward the CP-invariant 
vacuum $\theta_{\cancel{\tt CP}}=0$ by discharging ${\cal H}$ very prolifically, thanks to (\ref{boundrate}). 
This will continue until $| F_{\mu\nu\lambda\sigma} | <{2\pi^2 \sqrt{\cal X}}  {\cal Q}$. 
Once $| F_{\mu\nu\lambda\sigma} |$ is so small, the net CP-violating phase will be reduced to
\be
\theta_{\cancel{\tt CP}} \la  \frac{\cal Q}{\sqrt{\cal X}} \le 10^{-10} \, ,
\label{finalphase}
\ee
and the nucleations will cease. 

The charges and tensions in the range of $keV$ are small enough to 
ensure that the decay rate (\ref{nucrate}) is fast compared
to the age of the universe near and below the QCD scale, as seen in (\ref{boundrate}). The nucleations  
rapidly minimize CP-breaking effects throughout the universe\footnote{Interestingly, 
the discharge of $\theta_{\cancel{\tt CP}}$ toward zero also reduces the vacuum 
energy contribution from the QCD confinement phase transition. 
To study this in more detail,
one must go beyond the decoupling limit of gravity adopted 
in this paper, and so here we ignore it for the time being.}. 
This dynamics is very similar to how the axion relaxes
$\theta_{\cancel{\tt CP}}$. The main difference is that in our case, the relaxation occurs in small discrete steps 
instead of continuously. However, since this is happening during radiation domination,  
the minimal $\theta_{\cancel{\tt CP}}$ regions will percolate. Their membrane mantles 
will expand at the speed of light, and will collide with each other in a time shorter than the
Hubble time since the nucleation is so prolific (\ref{boundrate}). 

Upon a collision, the membranes would burn out by production of 
strongly interacting particles, i.e. predominantly pions. 
To infer this, using (\ref{cantradcharged}), we see that below the chiral symmetry breaking, we have 
\be
\int d^4x {\cal L}_{F + {\cal H}} \ni \frac{1}{2\pi^2 \sqrt{\cal X}} \int d^4x \, 
 {\cal H} \epsilon^{\mu\nu\lambda\sigma} F_{\mu\nu\lambda\sigma} \sim 
\frac{12}{\pi^2 \sqrt{\cal X}} \int d^4x \,  {\cal H} \partial_\mu K^\mu \, ,
\label{pions}
\ee
where in the last step we used the perturbative definition of the top form. 
After a chiral transformation the Chern-Simons
current will shift, on shell, to $\partial K \rightarrow \partial K + f_\pi \partial^2 {\cal P}$, 
with ${\cal P}$ denoting the pion field. 
Substituting into (\ref{pions}), integrating by parts and recalling that $\partial {\cal H} = 0$ away 
from membranes, and  $\Delta  {\cal H} = {\cal Q}$ on a membrane,
\be
\frac{12  f_\pi}{\pi^2 \sqrt{\cal X}} \int d^4x \,  {\cal H} \partial^2 {\cal P} 
\sim 10^{-11}  f_\pi \int_{R \times S^2} d^3 \xi \, n \cdot \partial {\cal P} \, ,
\label{pionsmem}
\ee
where $n$ is the outward normal to the membrane, and we substituted the membrane charge value.
At nucleation, the membranes are 
very small, $r_0 H_{\tt QCD} \sim 10^{-22} - 10^{-12}$, moving out at the speed of light. By the time they 
reach the size of the horizon at the QCD scale, they will have propagated through the 
background FRW for a time $\sim 1/H_{\tt QCD}$ which is at least a factor of $10^{12}$ larger than their initial size. 
The emission rate accumulates over a Hubble time, and the energy transfer of the membrane walls 
can be completed, compensating the suppression of $10^{-11}$ in Eq. (\ref{pionsmem}). 

To summarize, in this work we presented a mechanism 
of cosmological discharge of strong CP-violating phase by rapid membrane
production. These processes start right after chiral symmetry breaking. They look like fast 
``evaporation" of false vacua. 
When the membrane tension and charge are in the $keV$ range the discharges are fast enough 
to complete the relaxation of  $\theta_{\cancel{\tt CP}}$, and so approximately restore CP, well before BBN.
As a consequence the resulting cosmology passes the observational bounds. A small membrane charge 
ensures that the discharge will be refined enough to yield the terminal $\theta_{\cancel{\tt CP}}$ below
$10^{-10}$, and a small membrane tension then makes the discharges fast, achieving  
$\theta_{\cancel{\tt CP}} \la 10^{-10}$ quickly. 

The mechanism we proposed here must begin right around chiral 
symmetry breaking, discharging $\theta_{\cancel{\tt CP}}$ 
as soon as $\theta_{\cancel{\tt CP}}$ degeneracy is lifted. Many individual discharge processes are needed.
Hence to complete the transition to terminal $\theta_{\cancel{\tt CP}}$ we need many bubbles to form,  which 
complicates the description of the process. Another question concerns the origin of the scale of charges and tensions.
Small charges might appear  if there is kinetic mixing, imitating 
how the millicharge particles arise in field theory \cite{Holdom:1985ag}. Alternatively, they might come in models with
UV completions with large extra dimensions, where $4D$ charges 
include a factor of compact internal space volume 
\cite{Feng:2000if}. 

Small tensions could come from wrapping $p$-branes on 
shrinking cycles in internal dimensions \cite{Feng:2000if}. Since their scale is picked to make transitions fast, an 
alternative approach might be  faster false vacua decay rates, which could be done in a variety of ways
\cite{Garriga:2003gv,Brown:2015kgj,Brown:2010bc,Brown:2010mg,Easther:2009ft,Kleban:2011cs}. 
Another possibility might also
be using heavier axions, which should sit in their potential minima at chiral symmetry breaking, 
being effectively rendered immobile by their large mass and so decoupled, but where potentials
have a large number of vacua which could have enhanced tunneling rates 
\cite{Bachlechner:2017zpb}.
In any case, an alternative mechanism to relax $\theta_{\cancel{\tt CP}}$ 
warrants a closer look. It has been known for some time
that in theories with strong coupling in the IR topological defects may play a 
role of fundamental excitations. We advocate
that it is interesting to explore how their dynamics could help with 
understanding deeper problems of fundamental physics
of point-like particles.

\vskip.3cm

{\bf Acknowledgments}: The author thanks A. Lawrence and 
A. Westphal for useful discussions, and G. Dvali for comments. The author is grateful to KITP UCSB for kind
hospitality in the course of this work. The research reported here 
was supported in part by the DOE Grant DE-SC0009999. This research
was also supported in part by grant NSF PHY-2309135 and the Gordon and Betty Moore
Foundation Grant No. 2919.02 to the Kavli Institute for Theoretical Physics (KITP).

\end{document}